\documentclass[a4paper]{jpconf}
\usepackage{amsfonts,amsmath,amssymb,amsthm,wasysym,latexsym}
\usepackage{mathtools,bbold}
\usepackage{float}
\usepackage{graphicx}

\usepackage{fancyhdr}				
\begin{document}

\restylefloat{table}

\newcommand{\defeq}{\vcentcolon=}
\newcommand{\eqdef}{=\vcentcolon}
\renewcommand{\r}[1]{\mathrm{#1}}
\newcommand{\ts}[1]{{\mathrm{\scriptscriptstyle{#1}}}}
\newcommand{\sset}[1]{ \{#1\} }
\newcommand{\braket}[1]{\left\langle{#1}\right\rangle}
\newcommand{\bra}[1]{\left\langle{#1}\right|}
\newcommand{\ket}[1]{\left|{#1}\right\rangle}
\newcommand{\ketbra}[1]{\ket{#1}\!\bra{#1}}
\newcommand{\av}[1]{\left[{#1}\right]_{\ts{av}}}
\newcommand{\aT}[1]{\braket{#1}_{\ts{T}}}
\newcommand{\op}[1]{\hat{\mathrm{#1}}}
\newcommand{\vect}[1]{\boldsymbol{\mathrm{#1}}}
\newcommand{\prima}{^\prime}
\newcommand{\choice}[1]{\left\{\begin{array}{cl}#1\end{array}\right.}
\newcommand{\abs}[1]{\left|#1\right|}
\newcommand{\kB}{k_{\ts{B}}}
\newcommand{\kBT}{\kB T}
\newcommand{\ZZ}{\mathbb{Z}}
\newcommand{\cC}{\mathcal{C}}
\newcommand{\cE}{\mathcal{E}}
\newcommand{\cH}{\mathcal{H}}
\newcommand{\cO}{\mathcal{O}}
\newcommand{\cP}{\mathcal{P}}
\newcommand{\cR}{\mathcal{R}}
\newcommand{\cS}{\mathcal{S}}
\newcommand{\cZ}{\mathcal{Z}}
\newcommand{\cG}{\mathcal{G}}
\newcommand{\X}{\op{X}}
\newcommand{\Y}{\op{Y}}
\newcommand{\Z}{\op{Z}}
\newcommand{\I}{\op{\mathbb{1}}}
\newcommand{\dd}{\mathrm{d}}
\newcommand{\tetra}{\triangle}
\newcommand{\CNOT}{CNOT}
\newcommand{\hadamard}{H}
\newcommand{\phasegate}{K}
\newcommand{\hilbert}{\cH}
\newcommand{\code}{\cC}
\newcommand{\stabilizer}{\cS}
\renewcommand{\check}{\op{S}}
\newcommand{\projector}{\op{P}}
\newcommand{\checkX}{\check^{\ts{X}}}
\newcommand{\checkY}{\check^{\ts{Y}}}
\newcommand{\checkZ}{\check^{\ts{Z}}}
\newcommand{\checkw}{\check^{w}}
\newcommand{\sigmaX}{\op{\sigma}^{\ts{X}}}
\newcommand{\sigmaY}{\op{\sigma}^{\ts{Y}}}
\newcommand{\sigmaZ}{\op{\sigma}^{\ts{Z}}}
\newcommand{\sigmaw}{\op{\sigma}^{w}}
\newcommand{\spinX}{\spin^{\ts{X}}}
\newcommand{\spinY}{\spin^{\ts{Y}}}
\newcommand{\spinZ}{\spin^{\ts{Z}}}
\newcommand{\spinS}{\spin^{\ts{s}}}
\newcommand{\spinT}{\spin^{\ts{t}}}
\newcommand{\spinpZ}{\spin'^{\ts{Z}}}
\newcommand{\spinZZ}{\spin^{\ts{ZZ}}}
\newcommand{\spinw}{\spin^{w}}
\newcommand{\tauX}{\tau^{\ts{X}}}
\newcommand{\tauY}{\tau^{\ts{Y}}}
\newcommand{\tauZ}{\tau^{\ts{Z}}}
\newcommand{\tauw}{\tau^{w}}
\newcommand{\centralizer}{C}
\newcommand{\normalizer}{N}
\newcommand{\pauligroup}{{\cP_n}}
\newcommand{\gaugegroup}{\cG}
\newcommand{\hamiltonian}{\mathcal{H}}
\newcommand{\partition}{\cZ}
\newcommand{\spin}{s}
\newcommand{\state}{\psi}
\newcommand{\Tc}{T_{\ts{c}}}
\newcommand{\pc}{p_{\ts{c}}}
\newcommand{\obs}{\cO}
\newcommand{\channel}{\cE}
\newcommand{\error}{E}
\newcommand{\echain}{E}
\newcommand{\cycle}{C}
\newcommand{\prob}{\cP}
\newcommand{\eqprob}{\prob_{\ts{eq}}}
\newcommand{\recovery}{\cR}
\newcommand{\sublattice}{\lambda}
\newcommand{\bighexagon}{{\mbox{\LARGE\hexagon}}}
\newcommand{\smallhex}{{\mbox{\tiny\hexagon}}}
\newcommand{\smallsquare}{{\scriptscriptstyle{\square}}}
\newcommand{\lattice}{\Lambda}

\title{Stability of topologically-protected quantum computing proposals
as seen through spin glasses}

\author{H G  Katzgraber}
\address{Department of Physics \& Astronomy, Texas A\&M University, 
College Station, Texas 77843}
\address{Materials Science \& Engineering, Texas A\&M University, 
College Station, Texas 77843}
\ead{hgk@tamu.edu}

\author{R S Andrist}
\address{Theoretische Physik, ETH Zurich, CH-8093 Zurich, Switzerland}
\ead{andrist@phys.ethz.ch}

\begin{abstract}

Sensitivity to noise makes most of the current quantum computing schemes
prone to error and nonscalable, allowing only for small
proof-of-principle devices. Topologically-protected quantum computing
aims at solving this problem by encoding quantum bits and gates in
topological properties of the hardware medium that are immune to noise
that does not impact the entire system at once. There are different
approaches to achieve topological stability or active error correction,
ranging from quasiparticle braidings to spin models and topological
colour codes. The stability of these proposals against noise can be
quantified by their error threshold. This figure of merit can be
computed by mapping the problem onto complex statistical-mechanical
spin-glass models with local disorder on nontrival lattices that can
have many-body interactions and are sometimes described by lattice gauge
theories.  The error threshold for a given source of error then
represents the point in the temperature-disorder phase diagram where a
stable symmetry-broken phase vanishes. An overview of the techniques
used to estimate the error thresholds is given, as well as a summary of
recent results on the stability of different topologically-protected
quantum computing schemes to different error sources.

\end{abstract}

\section{Introduction}
\label{sec:intro}

Topological error-correction codes represent an appealing alternative to
traditional \cite{shor:95,steane:96,knill:97} quantum error-correction
approaches. Interaction with the environment is unavoidable in quantum
systems and, as such, efficient approaches that are robust to errors
represent the holy grail of this field of research. Traditional
approaches to error correction require, in general, many additional
quantum bits, thus potentially making the system more prone to failures.
However, topologically-protected quantum computation presents a robust
and scalable approach to quantum error correction: The quantum bits are
encoded in delocalized, topological properties of a system, i.e.,
logical qubits are encoded using physical qubits on a nontrivial surface
\cite{dennis:02}. Topological quantum error-correcting codes
\cite{kitaev:03,bombin:06,bombin:07,bombin:10,bravyi:10,haah:11} are
thus instances of stabilizer codes \cite{gottesman:96,calderbank:96}, in
which errors are diagnosed by measuring check operators (stabilizers).
In topological codes these check operators are local, thus keeping
things simple. The ultimate goal is not only to achieve good quantum
memories, but also to reliably perform computations.

The first topological quantum error-correction code was the Kitaev toric
code \cite{kitaev:03}. Other proposals followed, such as colour codes
\cite{bombin:06,bombin:07,bombin:10}, as well as stabilizer subsystem
codes \cite{poulin:05,bacon:06}. Interestingly, topological quantum
error correction has a beautiful and deep connection to classical
spin-glass models \cite{binder:86} and lattice gauge theories
\cite{nishimori:01,dennis:02}: When computing the error stability of the
quantum code to different error sources (e.g., qubit flips, measurement
errors, depolarization, etc.) the problem maps onto disordered
statistical-mechanical Ising spin models on nontrivial topologies with
$N$-body interactions. Furthermore, for some specific error sources, the
problem maps onto novel lattice gauge theories with disorder.

This paper outlines the close relationship between several realizations
of these error-correction strategies based on topology to classical
statistical-mechanical spin models. The involved mapping associates
faulty physical qubits with ``defective'' spin-spin interactions, as
well as imperfections in the error-correction process with flipped
domains. Thus a disordered spin state---characterized by system-spanning
domain walls---can be identified with the proliferation of minor errors
in the quantum memory. As a result, the mapping can be used to calculate
the error threshold of the original quantum proposal
\cite{dennis:02,raussendorf:07,katzgraber:09c,barrett:10,duclos:10,wang:11,landahl:11,bombin:12}:
If the spin system remains ordered, we know that error correction is
feasible for a given error source and an underlying quantum setup.
Because the quantum problem maps onto a disordered Ising spin-glass-like
Hamiltonian \cite{binder:86}, no analytical solutions exist. As such,
the computation of the error thresholds strongly depends on numerical
approaches.

Different methods to compute the error thresholds exist, ranging from
zero-temperature approaches that use exact matching algorithms (see, for
example, \cite{wang:11}), to duality methods
\cite{ohzeki:08,ohzeki:09,ohzeki:09a,ohzeki:11}. Unfortunately, the
former only delivers an upper bound, while the latter is restricted to
problems defined on planar graphs. A generic, albeit
numerically-intensive approach that allows one to compute the error
threshold for any error source (i.e., for any type of $N$-body Ising
spin glass on any topology) is given via Monte Carlo simulations
\cite{landau:00,katzgraber:09e}.

In section~\ref{sec:mapping} we outline the quantum--to--statistical
mapping for the case of the toric code \cite{dennis:02,kitaev:03}. In
this particular case, computing the error tolerance of quantum error
correction due to qubit flips maps onto a two-dimensional random-bond
Ising model \cite{binder:86} with additional requirements imposed on the
random couplings. The Monte Carlo methods used are described in
section~\ref{sec:methods}. Beyond the toric code, an equivalent mapping is
also possible for more involved error-correction schemes, more realistic
error sources, as well as under the assumption of an imperfect quantum
measurement apparatus.  Section~\ref{sec:results} summarizes our results
for different topologically-protected quantum computing proposals to
different error sources and a summary is presented in
section~\ref{sec:summary}.

Besides providing new and interesting classical statistical-mechanical
models to study, the results accentuate the feasibility of topological
error correction and raise hopes in the endeavor towards efficient and
reliable quantum computation.

\section{Mapping topological qubits onto spin glasses: Example of the
toric code}
\label{sec:mapping}

During the error-correction process, different errors can have the same
error syndrome \cite{kitaev:03,dennis:02} and we cannot determine which
error occurred. The best way to proceed is by classifying errors into
classes with the same effect on the system, i.e., errors that share the
same error-correction procedure. Once the classification is complete, we
correct for the most probable error class. Successful error correction
then amounts to the probability of identifying the correct error class.

In topological error-correction codes, this is achieved by measuring
local stabilizer operators. These are projective quantum measurements
acting on multiple neighboring qubits in order to determine, for example
in the case of qubit flip errors, their flip-parity. The actual quantum
operators are chosen carefully to allow for the detection of a flipped
qubit in a group without measuring (and thus affecting) the encoded
quantum information. Due to this limitation, the stabilizer measurements
can only provide some information about the location of errors, which is
then used to determine the most probable error class.

\begin{figure}
    \begin{minipage}[b]{4in}
	\includegraphics[width=4in]{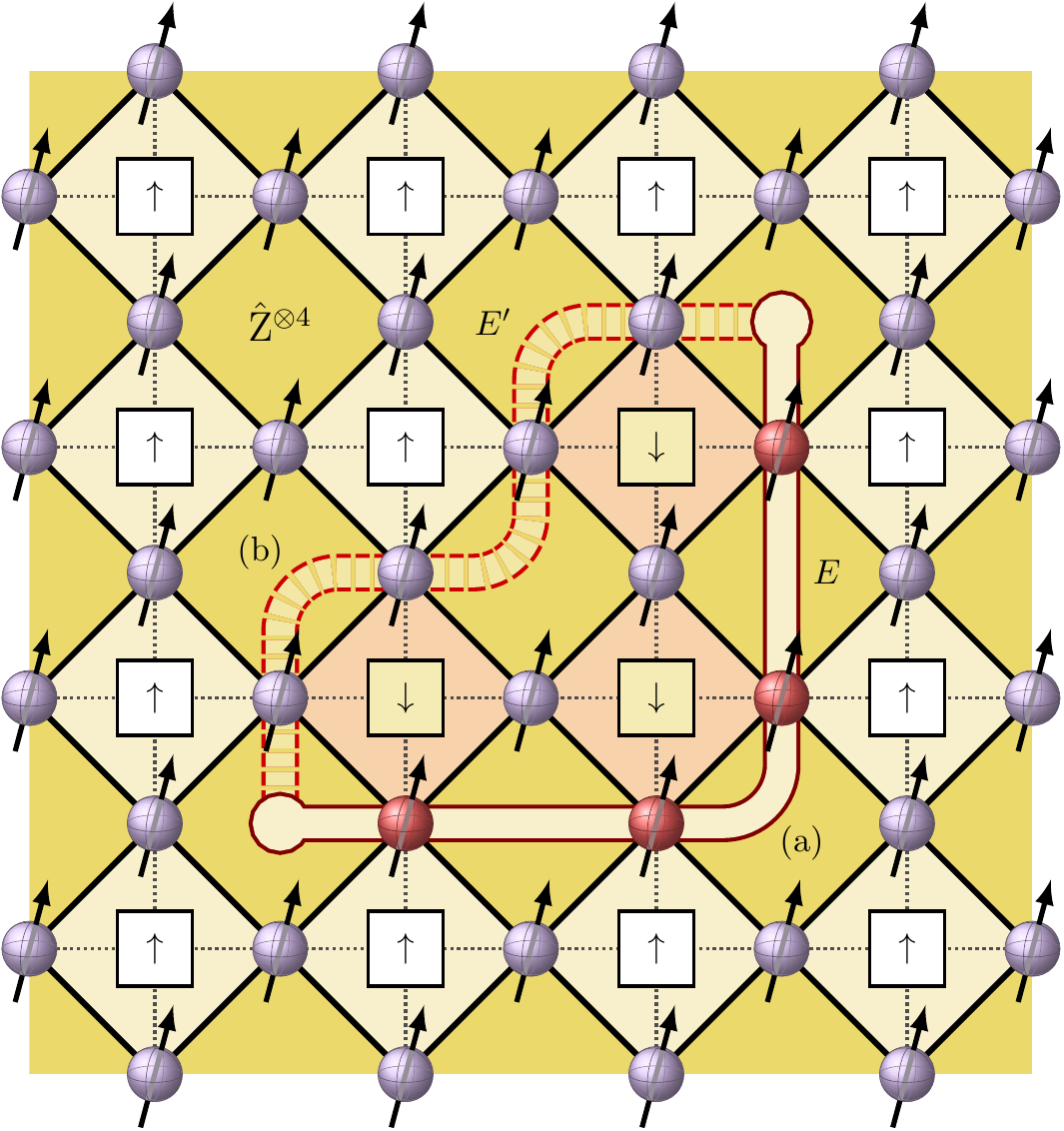}
    \end{minipage} \hfill \begin{minipage}[b]{2in}
	\caption{
	    For the toric code, qubits are arranged on a square lattice,
	    with stabilizer operators acting on plaquettes of four
	    qubits. The figure shows the spin placement to sample
	    chains from an error class $\bar\echain$: (a) A reference
	    error chain $\echain$ defines the error class; the
	    interactions are anti-ferromagnetic along this chain.  (b) A
	    chain which differs from the reference chain by a product of
	    spin-plaquettes: Together with the reference chain it forms
	    the boundary $\echain+\echain'$ of a flipped domain of
	    spins. The terms along $\echain'$ are excited, while the
	    ones along $\echain$ are now satisfied. Thus the Boltzmann
	    weight for this configuration is proportional to
	    $(p/(1-p))^{|\echain'|-|\echain'|}$.
	} \label{fig:toricspins}
    \end{minipage}
\end{figure}

The Kitaev proposal for the toric code arranges qubits on a square
lattice with stabilizer operators of the form $\Z^{\otimes4}$ ($\Z$ a
Pauli operator) and $\X^{\otimes4}$ ($\X$ a Pauli operator) acting on
the four qubits around each plaquette. When only qubit-flip errors are
considered, it is sufficient to treat only stabilizers of type
$\Z^{\otimes4}$ which are placed on the dark tiles of the checkerboard
decomposition, see figure~\ref{fig:toricspins}.
The measurement outcome of each stabilizer applied to its four
surrounding qubits is $\pm1$, depending on the parity of the number of
flipped qubits. These parity counts are not sufficient to locate the
exact position of the qubit errors, but for sufficiently low error rates
it is still possible to recover using this partial information (see 
\cite{kitaev:03} for details).  This can be achieved by
interpreting sets of neighboring errors as chains and classifying them
into error classes with the same effect on the encoded information.
During the error-correction process, all stabilizer operators are
measured and the resulting error syndrome represents the end points of
error chains. We refer to these sets of errors as chains, because two
adjacent errors cause a stabilizer to signal even flip-parity -- only
the end points of the chain are actually detected. This information is
still highly ambiguous in terms of the actual error chain $\echain$,
where the errors occurred. Fortunately, we do not need to know where
exactly the error occurred: Successful error correction amounts to
applying the error-correction procedure for an error from the correct
error class, i.e., such that no system spanning loop is introduced. The
question of whether error recovery is feasible therefore is determined
by the probability of identifying the correct error class. This
likelihood is what can be calculated through the mapping to classical
spin glasses.

For a constant qubit error rate $p$, the probability $\prob$ for a
specific error chain $\echain$ is determined by the number of faulty
qubits $\abs{\echain}$:
\begin{equation}
    \prob(\echain) = (1-p)^{N-\abs{\echain}}(p)^{\abs{\echain}}\,,
    \label{eq:toric:chain:prob}
\end{equation}
where $N$ is the number of qubits in the setup.
Equivalently, we can describe this error chain with Boolean values
$n_{\ell}^\echain\in\sset{0,1}$ for each qubit $\ell$, describing whether an
error occurred. The probability in \eqref{eq:toric:chain:prob} can
then be written as
\begin{equation}
    \prob(\echain) =
    \prod_{\ell}\,(1-p)^{N-n_{\ell}^\echain}(p)^{n_{\ell}^\echain}
    =(1-p)^{N}\prod_{\ell}\,
    \left(\frac{p}{1-p}\right)^{n_{\ell}^\echain}\,,
    \label{eq:toric:chain:prob2}
\end{equation}
where the product is over all qubits $\ell$. Because the stabilizer
measurements only yield the boundary of the error chain, there are many
other error chains $\echain'$ that are compatible with the same error
syndrome. If two chains $\echain$ and $\echain'$ share the same
boundary, then they can only differ by a set of cycles $\cycle$, which
have no boundary. The relative probability of $\echain'=\echain+\cycle$,
compared to $\echain$, depends on the relative number of faulty qubits,
which increases for every qubit in $\cycle\setminus\echain$ but
decreases for qubits in $\cycle\cap\echain$. Therefore, using analogous
Boolean descriptors for the chain, $n_{\ell}^\cycle$, we can write the
relative probability $\prob(\echain')/\prob(\echain)$ as:
\begin{equation}
    \frac{\prob(\echain')}{\prob(\echain)} = 
        \prod_{\ell}\left(\frac{p}{1-p}\right)^{
            n_{\ell}^\cycle(1-2n_{\ell}^\echain)
        }
    \propto \prod_{\ell}
    \exp[\beta J\tau_{\ell}(\underbrace{1-2n_{\ell}^\cycle}_{u_{\ell}})]\,.
    \label{eq:toric:chain:prob:rel}
\end{equation}
The newly-defined variable $u_{\ell}\in\sset{\pm1}$ is negative for all
links in $\cycle$ and we have introduced carefully-chosen coupling
constants $\tau_{\ell}\in\sset{\pm1}$ and a factor $\beta J$ such that
\begin{equation}
    e^{-2\beta J\tau_{\ell}} = [p/(1-p)]^{1-2n_{\ell}^\echain}\,.
    \label{eq:toric:nishimori2}
\end{equation}
Note that the sign of $\tau_{\ell}$ is dictated by the presence of an
error in the reference chain $\echain$, and $J$ is related to the error
probability via the Nishimori condition~\cite{nishimori:81}:
\begin{equation}
    -2\beta J = \ln\left(\frac{p}{1-p}\right)\,.
    \label{eq:toric:nishimori}
\end{equation}
The constraint for $\cycle$ to be cyclic (no boundary) imposes the
additional requirement that the number of adjacent faulty qubits
$\ell\in\cycle$ must be even for every plaquette. One way to satisfy
this condition is to introduce Ising variables $\spin_i\in\sset{\pm1}$
for each plaquette of the \emph{opposite} colour.  That is, each spin
represents an elementary cycle around a plaquette and larger cycles are
formed by combining several of these elementary loops.  For any choice
of the spin variables $\spin_i$, the variables $u_\ell =
\spin_i\spin_j$, with $\ell$ the edge between plaquettes $i$ and $j$,
describes such a cyclic set $\cycle$ (see figure~\ref{fig:toricspins}).

We have therefore found that the spin configurations enumerate all error
chains $\echain'$ that differ from the reference chain $\echain$ by a
tileable set of cycles. With the Nishimori relation
\eqref{eq:toric:nishimori} their Boltzmann weight is also
proportional to the probability of the respective error chain.
Therefore, it is possible to sample the fluctuations of error chains
within the same error class by sampling configuration from the classical
statistical model described by the partition function
\begin{equation}
    \partition_{\sset{\tau_{ij}}} = \sum_{\sset{\spin}}
        \exp \big( -\beta J\sum_{\braket{i,j}} \tau_{ij}\spin_i\spin_j \big)\,.
    \label{eq:codes:partition}
\end{equation}
Here $J$ is dictated by the Nishimori condition
\eqref{eq:toric:nishimori} and $\tau_{ij}$ are quenched,
disordered interactions which are negative if the associated qubit is
faulty in the reference chain. Because the mapping identifies error
chains with domain walls and their difference with a flipped patch of
spins, we can identify the ordered state with the scenario where error
chain fluctuations are small and correct error classification is
feasible. And while this sampling does not implicitly consider
homologically nontrivial cycles, we can interpret percolating domain
walls as error fluctuations which are too strong to reliably distinguish
cycles of different homology.

Because $\beta J$ and $p$ can only be related on the Nishimori line for
the mapping between the quantum problem and the statistical-mechanical
counterpart to work, we need to compute the point in the disorder $p$
and critical temperature $\Tc(p)$ plane where the Nishimori line
\eqref{eq:toric:nishimori} intersects the phase boundary between a
paramagnetic and a ferromagnetic phase, see
figure~\ref{fig:toric:phasediagram}.  This point, $\pc$ then corresponds
to the error threshold of the underlying topologically-protected quantum
computing proposal. For the case of the toric code with qubit flip
errors, the problem maps onto a two-dimensional random-bond Ising model
described by the Hamiltonian:
\begin{equation}
    \hamiltonian = -J\sum_{\langle i,j\rangle}
        \tau_{ij}\spin_i\spin_j\,,
    \label{eq:toric:hamiltonian}
\end{equation}
where $J$ is a global coupling constant chosen according to the
Nishimori condition \eqref{eq:toric:nishimori},
$\spin_{i}\in\sset{\pm1}$, the sum is over nearest neighbor pairs, and
the mapping requires $\tau_{ij}$ to be quenched bimodal random
interactions, distributed according to the error rate $p$:
\begin{equation}
    \prob(\tau_{ij}) = \choice{
        +1\; ; & 1-p\\
        -1\; ; & p \, .
    }
\end{equation}
For the toric code with qubit flip errors $\pc \approx 10.9$\%
\cite{honecker:01,merz:02,ohzeki:09,parisen:09}, i.e., as long as the
fraction of faulty physical qubits does not exceed $10.9$\%, errors can
be corrected. In the next section we outline the procedure used to
estimate different error thresholds using Monte Carlo methods.

\section{Estimating error thresholds using Monte Carlo methods}
\label{sec:methods}

The partition function found in the mapping for Kitaev's toric code can
be interpreted as a classical spin model where the different spin
configurations are weighted proportional to the likelihood of the error
chain they represent. The existence of such a relationship is
instrumental in understanding the fluctuations of error chains, because
it allows for the computation of the thermodynamic value of the error
threshold using tools and methods from the study of statistical physics
of disordered systems.

\subsection{Algorithms}

The simulations are done using the parallel tempering Monte Carlo method
\cite{swendsen:86,geyer:91,hukushima:96,marinari:96,katzgraber:06a}.
The method has proven to be a versatile ``workhorse'' in many fields
\cite{earl:05}. Similar to replica Monte Carlo \cite{swendsen:86},
simulated tempering \cite{marinari:92}, or extended ensemble methods
\cite{lyubartsev:92}, the algorithm aims to overcome free-energy
barriers in the free energy landscape by simulating several copies of a
given Hamiltonian at different temperatures. The system can thus escape
metastable states when wandering to higher temperatures and relax to
lower temperatures again in time scales several orders of magnitude
smaller than for a simple Monte Carlo simulation at one fixed
temperature. A delicate choosing of the individual temperatures is key
to ensure that the method works efficiently. See, for example,
\cite{katzgraber:06a}.

The classical Hamiltonians obtained when computing error thresholds in
topological quantum computing proposals all have quenched bond disorder,
i.e., a complex energy landscape with many metastable states. As such,
parallel tempering is the algorithm of choice when simulating these
systems, especially when temperatures are low and disorder high (for
example, close to the error threshold) where thermalization is
difficult.

Equilibration is typically tested in the following way: We study how the
results for different observables vary when the simulation time is
successively increased by factors of 2 (logarithmic binning). We require
that the last three results for all observables agree within error bars.

\subsection{Determination of the phase boundary}

To determine the error threshold we have to first determine the phase
boundary between an ordered and disordered phase. In most cases this
phase boundary can be determined by studying dimensionless functions of
local order parameters like the magnetization. Note, however, that some
models map onto lattice gauge theories where averages of all local order
parameters are zero. In these specific cases we use other approaches
outlined below.

\begin{figure}
    \begin{minipage}[b]{3in}
	\includegraphics[width=3in]{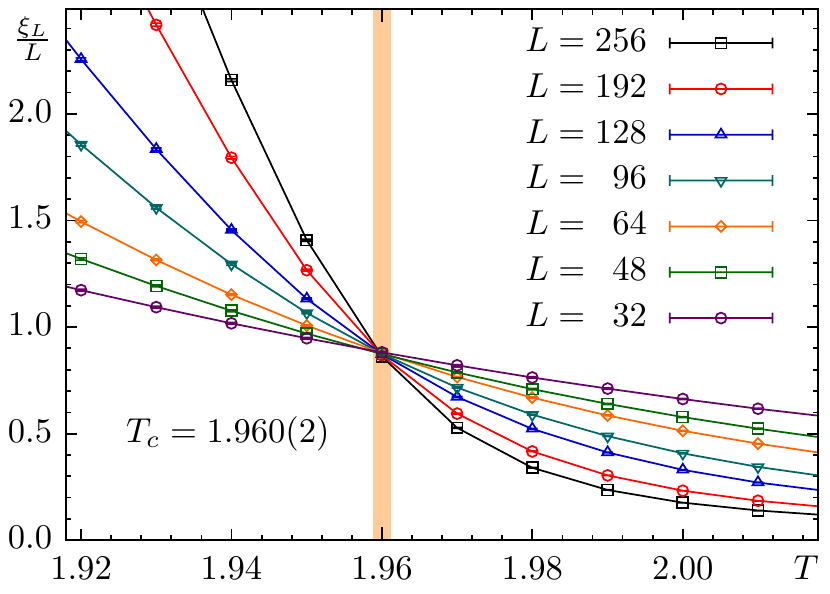}
    \end{minipage} \hfill \begin{minipage}[b]{3in}
	\caption{
	    Two-point finite-size correlation length $\xi_{L}/L$ as a
	    function of temperature for different system sizes $L$ for
	    the two-dimensional random-bond Ising model with a fraction
	    of $p = 0.04$ antiferromagnetic bonds. This corresponds to a
	    toric code with $4$\% faulty qubits (bit flip errors). The
	    data cross at $\Tc \approx 1.960(2)$, signaling a transition
	    (shaded area; the width represents the statistical error
	    bar). The data cross cleanly and show only small finite-size
	    corrections (these become stronger close to $\pc$).
	} \label{fig:toric:crossing}
    \end{minipage}
\end{figure}

As an example, for the case of the toric code, we study a transition
between a paramagnetic and a ferromagnetic phase.  The transition is
determined by a finite-size scaling of the dimensionless two-point
finite-size correlation length divided by the system size
\cite{cooper:82,palassini:99b,ballesteros:00,martin:02}.  We start by
determining the wave-vector-dependent susceptibility
\begin{equation}
\chi(k) =  \frac{1}{N} \sum_{ij}^N \langle S_iS_j \rangle_T \;
           e^{i{\bf k}\cdot({\bf R}_i - {\bf R}_j)} \; .
\label{eq:chik}
\end{equation}
Here, $\langle \cdots \rangle_T$ represents a thermal (Monte Carlo time)
average and ${\bf R}_i$ is the spatial location of the $N$ spins. The
correlation length is then given by
\begin{equation}
\xi_{L} = \frac{1}{2 \sin(k_{\rm min}/2)}
\sqrt{\frac{[\chi(k = 0)]_{\rm av}}{[\chi(k_{\rm min})]_{\rm av}}
- 1} ,
\label{eq:xiL}
\end{equation}
where $k_{\rm min} = (2 \pi / L,0)$ is the smallest nonzero wave vector
and $[\cdots]_{\rm av}$ represents an average over the different error
configurations (bond disorder). The finite-size correlation length
divided by the system size has the following simple finite-size scaling
form:
\begin{equation}
\xi_{L}/L \sim \widetilde{X}(L^{1/\nu}[T - \Tc]) ,
\label{eq:fss}
\end{equation}
where $\nu$ is a critical exponent and $\Tc$ represents the transition
temperature we need to construct the phase boundary.  Numerically,
finite systems of linear size $L$ are studied. In that case the function
$\xi_{L}/L$ is independent of $L$ whenever $T=\Tc$ because then the
argument of the function $\tilde{X}$ is zero.  In other words, if a
transition is present, the data cross at one point (up to corrections to
scaling).  This is illustrated in figure~\ref{fig:toric:crossing} for the
case of the toric code and $4$\% faulty qubits: Data for different
system sizes cross at $\Tc \approx 1.960(2)$ signaling a transition.
Because finite-size scaling corrections are typically small, one can use
the estimate of $\Tc$ obtained as a very good approximation to the
thermodynamic limit value.

\begin{figure}
    \begin{minipage}[b]{3in}
	\includegraphics[width=3in]{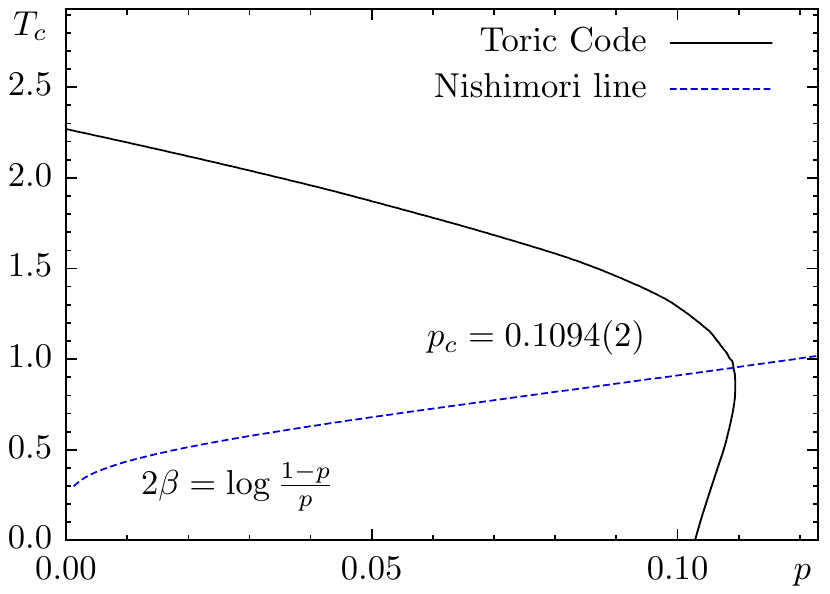}
    \end{minipage} \hfill \begin{minipage}[b]{3in}
	\caption{
	    Phase diagram for the two-dimensional random-bond Ising
	    model. $p$ represents the fraction of antiferromagnetic
	    bonds (fraction of flipped qubits) and $\Tc$ the transition
	    temperature. The dashed line is the Nishimori line. The
	    point where it intersects the phase boundary (solid line)
	    represents the error threshold which, in this case, is $\pc
	    \approx 0.1094(2)$
	    \cite{honecker:01,wang:03,merz:02,ohzeki:09,parisen:09}.
	    This means that errors can be corrected as long as no more
	    than $10.9$\% physical qubits have flipped.  Phase boundary
	    data taken from \cite{thomas:11d}.
	} \label{fig:toric:phasediagram}
    \end{minipage}
\end{figure}

The simulations must now be repeated for different fractions $p$ of
faulty qubits, i.e., for different fractions of
ferromagnetic-to-antiferromagnetic bonds between the classical Ising
spins. This then allows one to build a temperature--disorder phase
diagram as shown in figure~\ref{fig:toric:phasediagram} for the case of
the toric code. The error threshold then corresponds to the point where
the Nishimori line (dashed line in figure~\ref{fig:toric:phasediagram})
intersects the phase boundary (solid line in
figure~\ref{fig:toric:phasediagram}). In this particular case this occurs
for $\pc \approx 10.9$\%, i.e., as long as there are less than $10.9$\%
physical qubit flips, errors can be corrected in the quantum code.

\section{Results}
\label{sec:results}

We now summarize our results for different topological codes, as well as
different sources of error. Note that the mappings onto the
statistical-mechanical models are often complex and, for the sake of
brevity, we refer the reader to the individual manuscripts cited.

\subsection{Toric code with qubit-flip errors}

The toric code with qubit-flip errors has already been described in detail
in section~\ref{sec:mapping}. As stated before, the error-correction
process maps onto a two-dimensional random-bond Ising model:
\begin{equation}
    \hamiltonian = -J\sum_{i,j}
        \tau_{\langle ij\rangle }\spin_i\spin_j \, ,
	\;\;\;\;\;\;\;\;\;\;\;\;\;\;
	\prob(\tau_{ij}) = \choice{
        +1\; ; & 1-p\\
        -1\; ; & p
	}
    \label{eq:toric:hamiltonian2}
\end{equation}
where $J$ is a global coupling constant chosen according to the
Nishimori condition \eqref{eq:toric:nishimori},
$\spin_{i}\in\sset{\pm1}$, and the sum is over nearest neighbors.  The
error threshold for the toric code was computed by Dennis {\em et al.}
in \cite{dennis:02}, $\pc=0.1094(2)$, with a lower bound given by
Wang {\em et al.}~in \cite{wang:03} (the phase diagram is reentrant
\cite{wang:03,parisen:09,thomas:11d}). Note that more detailed estimates
of $\pc$ followed later \cite{honecker:01,merz:02,ohzeki:09,parisen:09}.
The associated phase boundary is shown in
figure~\ref{fig:toric:phasediagram}.

\subsection{Colour codes with qubit-flip errors}
\label{subsec:ccbf}

In 2006 Bomb\'in and Mart\'in-Delgado showed that the concept of
topologically-protected quantum bits could also be realized on trivalent
lattices with three-colourable faces. The \emph{topological colour codes}
they introduced~\cite{bombin:06} share similar properties to the toric
code: stabilizer generators are intrinsically local and the encoding is
in a topologically-protected code subspace. However, colour codes are
able to encode more qubits than the toric code and, for some specific
lattice structures, even gain additional computational capabilities.

In colour codes, qubits are arranged on a trivalent lattice (hexagonal or
square octagonal), such that each qubit contributes a term of the form
$\beta J\tau_{ijk}\spin_i\spin_j\spin_k$ in the mapping. In the case of
hexagonal lattices, the partition function takes the form
\cite{katzgraber:09c}
\begin{equation}
    \partition_{\sset{\tau_{ijk}}} = \sum_{\sset{\spin}} \exp\big(
        -\beta J\!\!\!\!\sum_{i,j,k\in\sset{\triangle}}\!\!\!
        \tau_{ijk}\spin_i\spin_j\spin_k\big)\,.
    \label{eq:color:partition}
\end{equation}
Equation \eqref{eq:color:partition} describes a disordered statistical
system with three-spin interaction for each plaquette. Note that the
spins $\spin_i$ defined for the mapping are located on the triangular
lattice which is \emph{dual} to the original hexagonal arrangement.
Every qubit corresponds to a triangle in the new lattice and dictates
the sign of the associated plaquette interaction via $\tau_{ijk}$.
Thus, the statistical-mechanical Hamiltonian for the system related to
colour codes [as described by \eqref{eq:color:partition}] is
given by
\begin{equation}
    \hamiltonian = -J\!\!\!\sum_{i,j,k\in\sset{\triangle}}\!\!
        \tau_{ijk}\spin_i\spin_j\spin_k\,,
    \label{eq:color:hamiltonian}
\end{equation}
where $J$ is a global coupling constant chosen according to the
Nishimori condition, $\spin_{i}\in\sset{\pm1}$, and the mapping requires
$\tau_{ijk}$ to satisfy
\begin{equation}
    \prob(\tau_{ijk}) = \choice{
        +1\; ; & 1-p\\
        -1\; ; & p
    }\,.
\end{equation}
Note that the disordered three-body Ising model on the triangular
lattice with $p = 0.5$ is NP-hard and therefore numerically difficult to
study \cite{thomas:11}.  We would like to emphasize that colour codes on
square-octagonal lattices are of particular interest, because, contrary
to both toric codes and colour codes on honeycomb lattices, they allow
for the transversal implementation of the whole Clifford group of
quantum gates.

\begin{figure}
    \begin{minipage}[b]{3in}
	\includegraphics[width=3in]{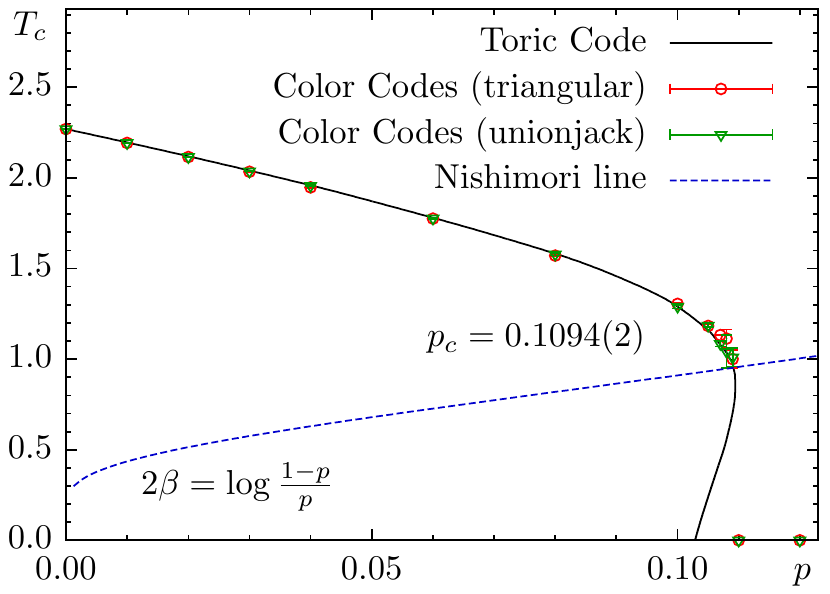}
    \end{minipage} \hfill \begin{minipage}[b]{3in}
	\caption{
	    Comparison of the $p\,$--$\,T_c$ phase diagram for the toric
	    code (solid black line) and the results for random
	    three-body Ising model on a triangular lattice (red), as
	    well as on a Union Jack lattice (green).  The error
	    threshold is indicated by the intersection of the phase
	    boundary with the Nishimori line (dashed blue) For $p > p_c
	    \approx 0.109$ the ferromagnetic order is lost.  Note the
	    agreement with the toric code. Data taken from
	    Refs.~\cite{katzgraber:09c}, \cite{katzgraber:10}, and
	    \cite{thomas:11d}.
	} \label{fig:color:phasediagram}
    \end{minipage}
\end{figure}

Figure \ref{fig:color:phasediagram} shows the $p$--$\Tc$ phase diagram
for colour codes on hexagonal (maps onto a triangular lattice; empty
circles), as well as square octagonal lattices (maps onto a Union Jack
lattice; empty triangles). In addition, the solid (black) line is the
phase boundary for the toric code.  Surprisingly, the phase boundaries
for all three models agree within statistical error bars, suggesting
that colour as well as toric codes share similar error thresholds $\pc$.
This is surprising, because the underlying statistical models have very
different symmetries and are in different universality classes. For
example, in the absence of randomness, the three-body Ising model on a
triangular lattice is in a different universality class than the
two-dimensional Ising model. Whereas for three-body Ising model on a
two-dimensional triangular lattice $\nu = \alpha = 2/3$
\cite{baxter:73}, for the two-dimensional Ising model $\nu = 1$ and
$\alpha = 0$. The disordered three-body Ising model on a triangular
lattice has not been studied before, therefore highlighting again the
fruitful relationship between quantum information theory and statistical
physics.

Finally, our results also show that the enhanced computing capabilities of
colour codes on the square-octagonal lattice do not come at the expense
of an increased susceptibility to noise.

\subsection{Depolarizing channel}

The effects of single-qubit operations can be decomposed into qubit
flips and phase flips, as well as a combination thereof; represented by
the three Pauli matrices $\X$, $\Z$, and $\Y$.  When describing
decoherence effects as a noisy channel, depolarizing noise is
characterized by equal probability for each type of error to occur,
i.e., $p_x = p_y = p_z \defeq p/3$. Note that the depolarizing channel
is more general than the bit-flip channel, because it allows for the
unified, correlated effect of all three basic types of errors
\cite{bombin:10,bombin:12}.

As in the previous mappings, in order to express the probability of an
error class in terms of a classical Boltzmann weight, we need to
associate with each elementary error loop a classical Ising spin.
However, because of the error correlations, we cannot treat errors of
different types independently any more. Instead, the resulting model
contains spins of different types, according to the types of stabilizers
used in the code. In fact, the mapping can be carried out in a very
general way that requires no assumptions on the individual error rates
or the actual quantum setup, see \cite{bombin:12}.  However, here
we merely provide a brief explanation of the resulting Hamiltonian for
the Toric code within the depolarizing channel.

\begin{figure}
    \begin{minipage}[b]{3in}
	\includegraphics[width=3in]{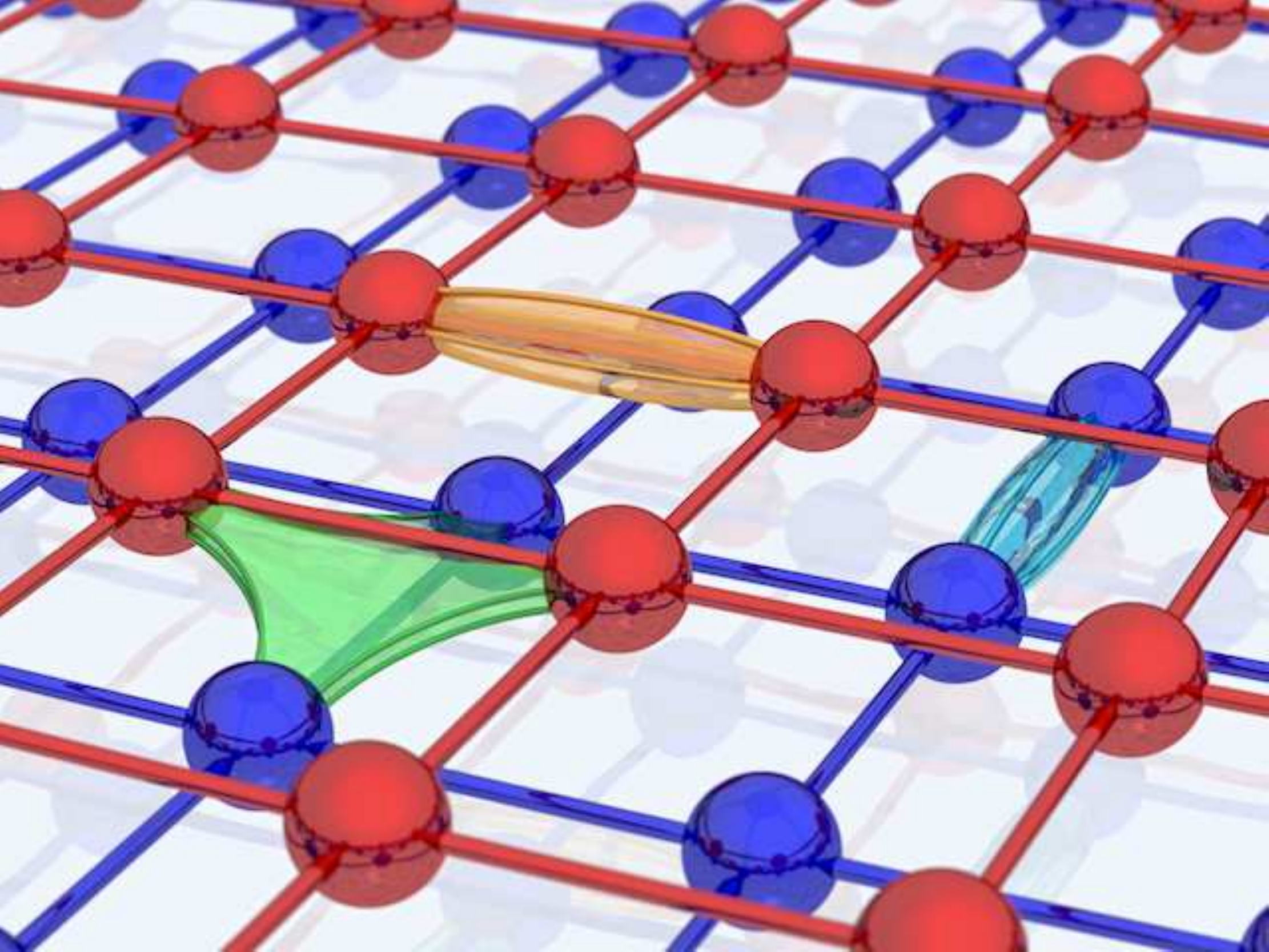} \caption{
	    When computing the stability of the toric code to
	    depolarization, the problem maps onto a classical
	    statistical Ising model on two stacked square lattices with
	    both two-body and four-body interactions.
	} \label{fig:depoltoriclattice}
    \end{minipage} \hfill \begin{minipage}[b]{3in}
	\includegraphics[width=3in]{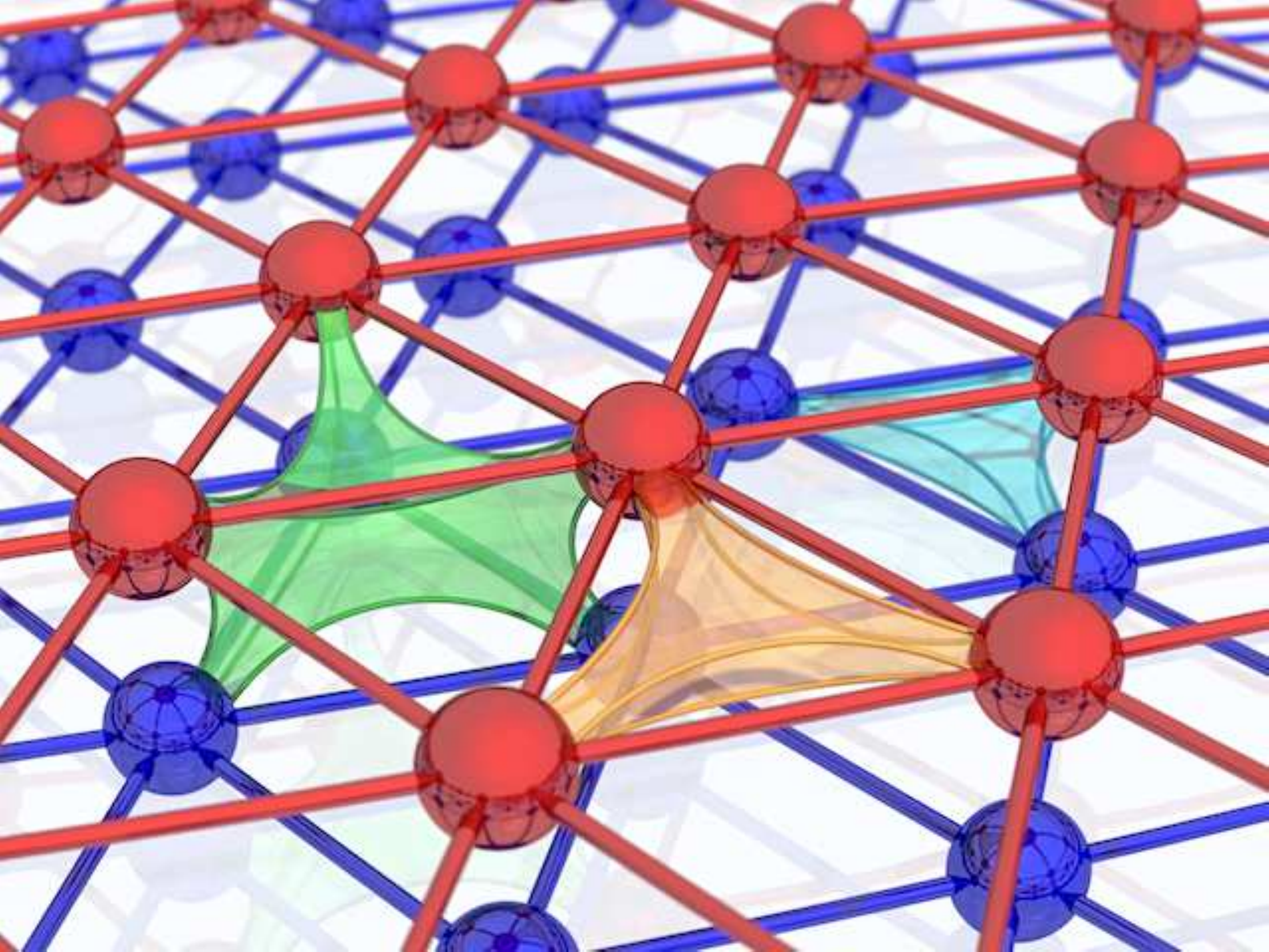} \caption{
	    For colour codes, the spins are arranged on triangular lattices.
	    In addition to the three-body interactions found for
	    qubit-flip errors, both layers are connected via six-body
	    interactions (light green).
	} \label{fig:depolcolorlattice}
    \end{minipage}
\end{figure}

In addition to the stabilizers of type $\Z^{\otimes4}$ (see
section~\ref{sec:mapping}), the toric code also places stabilizers of type
$\X^{\otimes4}$ on the remaining squares in the checkerboard
decomposition. These allow for the concurrent detection of possible
phase errors on the physical qubits. As a result, whenever a qubit
flips, this is signaled by adjacent $\Z$-stabilizers, whereas a qubit
attaining a phase error is identified by $\X$-stabilizers. Additionally,
a combined qubit flip and phase error affects both the neighboring $\Z$
and $\X$-stabilizers.  Therefore, the resulting Hamiltonian contains
three terms per qubit; one describing each of the aforementioned
scenarios:
\begin{equation}
        \hamiltonian = -J\sum_{ijk\ell} (
			\tau_{ij}^x s_i^zs_j^z + 
			\tau_{k\ell}^z s_k^xs_\ell^x + 
			\tau_{ijk\ell}^y s_i^zs_j^zs_k^xs_\ell^x
		)\,,
        \label{eq:depol:toric:hamiltonian}
\end{equation}
where the sum is over all qubits, the indices $i$, $j$, $k$ and $\ell$
denote the four affected elementary equivalences and the sign of
$\tau^w$ is dictated by whether the qubit has suffered an error of type
$w\in \{x, y, x\}$. This Hamiltonian describes a classical Ising model
that can be interpreted as two stacked square lattices which are shifted
by half a lattice spacing, see figure~\ref{fig:depoltoriclattice}. In
addition to the standard two-body interactions for the top and bottom
layers, the Hamiltonian also includes four-body terms (light green in
the figure) that introduce correlations between the layers.
Interestingly, toric codes under the depolarizing channel are related to
the eight-vertex model introduced by Sutherland \cite{sutherland:70}, as
well as Fan and Wu \cite{fan:70}, and later solved by
Baxter~\cite{baxter:71,baxter:72,baxter:82}.

For topological colour codes, qubits are arranged on trivalent lattices
(hexagonal or square-octagonal) and the problem then maps onto either a
triangular or Union Jack lattice (see section~\ref{subsec:ccbf}).  For the
depolarizing channel, an analogous mapping to the previous one relates
this quantum setup to a Hamiltonian of the form:
\begin{equation}
        \hamiltonian = -J\sum_{ijk} (
			\tau_{ijk}^x s_i^zs_j^zs_k^z + 
			\tau_{ijk}^z s_i^xs_j^xs_k^x + 
			\tau_{ijk}^y s_i^zs_j^zs_k^zs_i^xs_j^xs_k^x
		)\,.
        \label{eq:depol:toric:hamiltonian2}
\end{equation}
The details of this mapping are also contained in \cite{bombin:12}.
In addition to the in-plane three-body plaquette interactions that
appear already when studying individual qubit-flips (see
section~\ref{subsec:ccbf}), here additional six-body interactions add the
necessary correlations between the planes (see
figure~\ref{fig:depolcolorlattice}). The resulting model maps
to an eight-vertex model on a Kagom\'e lattice \cite{baxter:78}.

\begin{figure}
    \begin{minipage}[b]{3in}
	\includegraphics[width=3in]{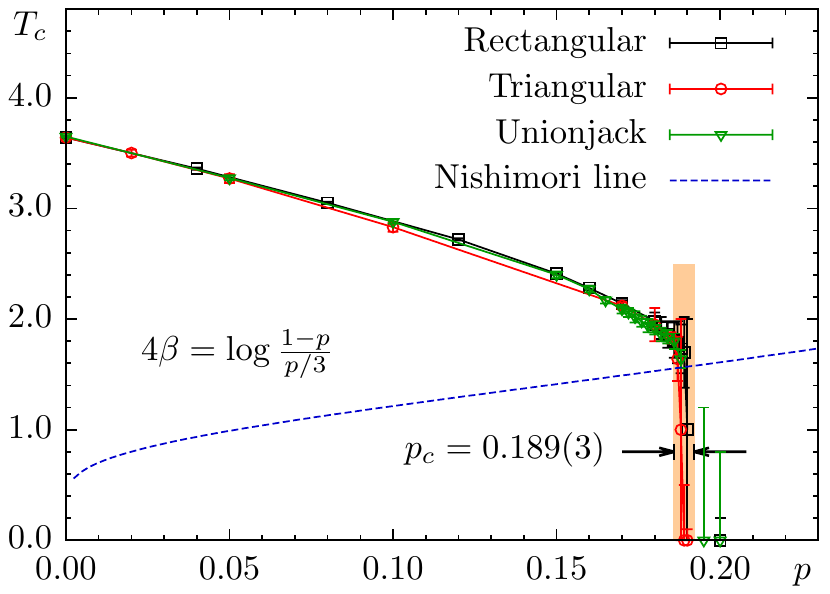}
    \end{minipage} \hfill \begin{minipage}[b]{3in}
	\caption{
	    Estimated phase boundary for the models related to the
	    depolarizing channel. The individual data sets represent the
	    toric code (black), as well as colour codes on triangular
	    (red) and Union Jack (green) lattices. The error threshold
	    $\pc = 0.189(3)$ corresponds to the point where the
	    Nishimori line (dashed, blue) intersects the phase boundary.
	    Remarkably, the phase boundaries for all three codes agree
	    within error bars.  Data taken from \cite{bombin:12}.
	} \label{fig:depolarize:phasediagram}
    \end{minipage}
\end{figure}

We perform simulations using the sublattice magnetization
\cite{bombin:12} to compute the susceptibility, and construct again the
$p\,$--$\,\Tc$ phase diagram for the toric code, as well as colour codes
on both hexagonal and square-octagonal lattices.  Note also, that the
Nishimori condition changes slightly in this particular context (see the
equation in figure~\ref{fig:depolarize:phasediagram} and
\cite{bombin:12}).  Interestingly, the phase boundaries for all
three error-correction models agree and we estimate conservatively an
error threshold of $\pc = 0.189(3)$ for all three models. A similar
study based on duality considerations \cite{bombin:12,ohzeki:08} yields
results that agree within error bars.  It is remarkable that the error
threshold to depolarization errors for different classes of topological
codes studied is larger than the threshold for expected uncorrelated
errors, $\pc'=(3/2)p_{c,\ts{flip}}\approx16.4\%$. This is encouraging
and shows that topological codes are more resilient to depolarization
than previously thought.  It also suggests that a detailed knowledge of
the correlations in the error source can allow for a more efficient,
custom-tailored code to be designed.

\subsection{Subsystem Codes}

All topological codes share the advantage that the quantum operators
involved in the error-correction procedure are local, thus rendering
actual physical realizations more feasible. However, in practice the
decoherence of quantum states is not the only source of errors: Both
syndrome measurement and qubit manipulations are delicate tasks and
should be kept as simple as possible to minimize errors. For the toric
code and topological colour codes, the check operators, while
local, still involve the combined measurement of $4$, $6$, or even
$8$ qubits at a time.

By using concepts from subsystem stabilizer codes, Bombin was able to
introduce a class of \emph{topological subsystem codes}~\cite{bombin:10}
that only requires pairs of neighboring qubits to be measured for
syndrome retrieval. This is achieved by designating some of the logical
qubits as ``gauge qubits'' where no information is stored. Due to the
resulting simplicity of the error-correction procedure, these are very
promising candidates for physical implementations.

\begin{figure}
    \begin{minipage}[b]{3in}
	\includegraphics[width=3in]{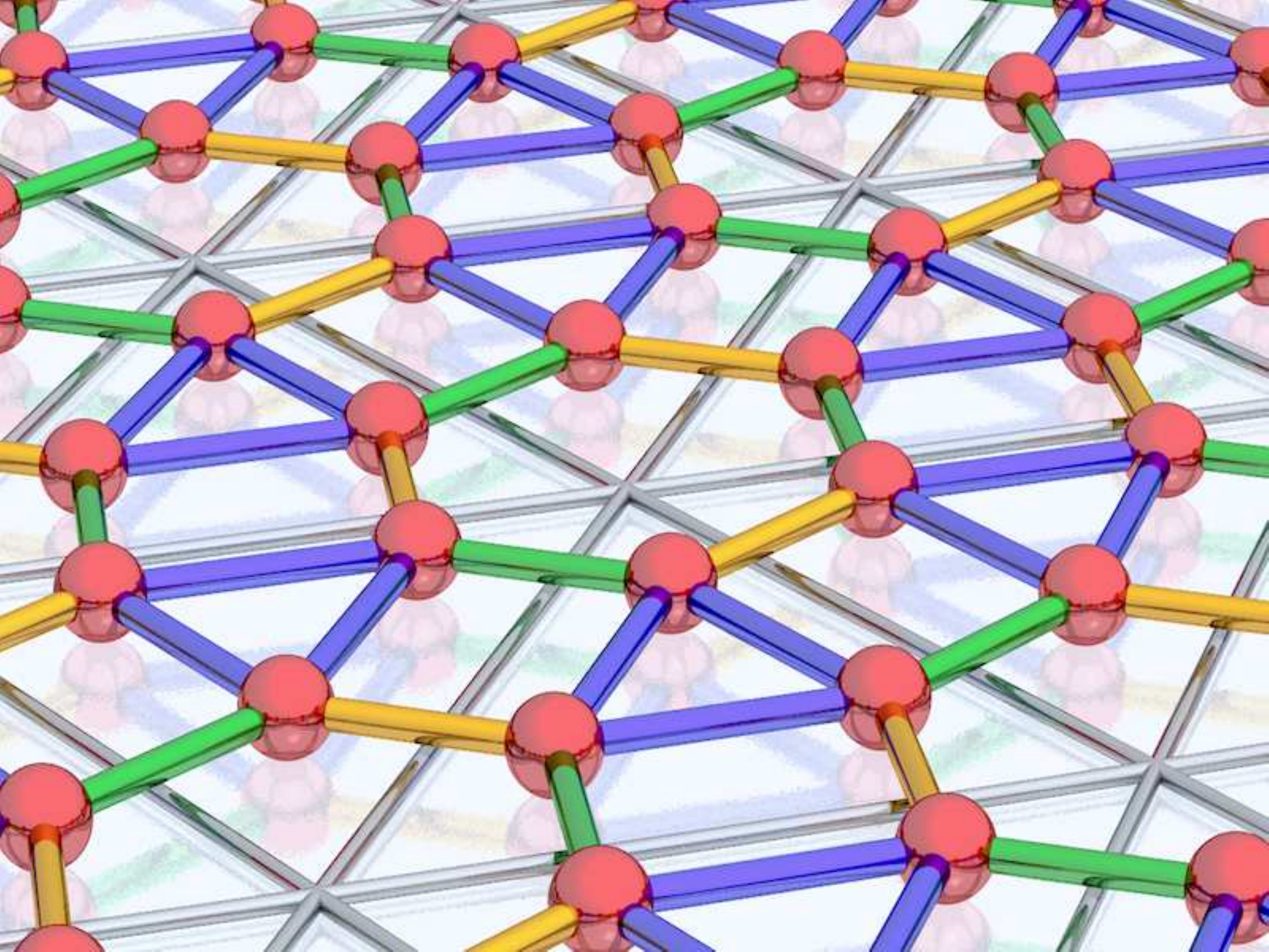}
    \end{minipage} \hfill \begin{minipage}[b]{3in}
	\caption{
	    Topological subsystem codes combine aspects of both
	    topological error-correction codes and subsystem codes.
	    Starting from a triangular lattice, the construction places
	    three qubits in each triangle with stabilizers acting on
	    pairs of them. Here the red spheres represent individual
	    qubits, interconnected with stabilizers of type $\X$
	    (yellow), $\Y$ (green) and $\Z$ (blue). The original
	    triangular lattice for the construction is shown in grey.
	    Despite only relying on two-qubit stabilizers, this setup is
	    able to preserve quantum information up to a threshold of
	    $5.5$\%.
	} \label{fig:subsystemlattice}
    \end{minipage}
\end{figure}

The generation and mapping of a quantum setup that incorporates all
these desired concepts is rather involved; we refer the reader to the
relevant papers in Refs.~\cite{bombin:10} and~\cite{andrist:12}. One
possible arrangement is shown in figure~\ref{fig:subsystemlattice},
consisting of qubits arranged in triangles, squares and hexagons with
stabilizer operators of different types connecting neighboring pairs. In
the mapping to a classical system, the setup then corresponds to a set
of Ising spins (one for each stabilizer) with interactions dictated by
how each stabilizer is affected by errors on adjacent qubits.  This
gives rise to a Hamiltonian of the general form
\begin{equation}
    \hamiltonian = -J
        \sum_\ell
        \sum_{w=x,y,z}
        \tau_\ell^w \prod_i \spin_i^{g_{i\ell}^w}\,,
    \label{eq:subsys:hamiltonian:general}
\end{equation}
where $\ell$ enumerates all qubit sites, $w$ the three error types and
$i$ iterates over all Ising spins, respectively. The exponent
$g_{i\ell}^w\in\sset{0,1}$ determines whether the stabilizer $i$ is
affected by an error of type $w$ on qubit $\ell$. Thus, for every qubit
$\ell$ and error type $w$, the Hamiltonian contains a term of the form
$-J\tau_\ell^w s_is_js_k$, where $J$ is a constant, $\tau_\ell^w$ is a
quenched random variable (representing a possible qubit error) and the
product contains all Ising spins corresponding to stabilizers affected
by such an error.

\begin{figure}
    \begin{minipage}[b]{3in}
	\includegraphics[width=3in]{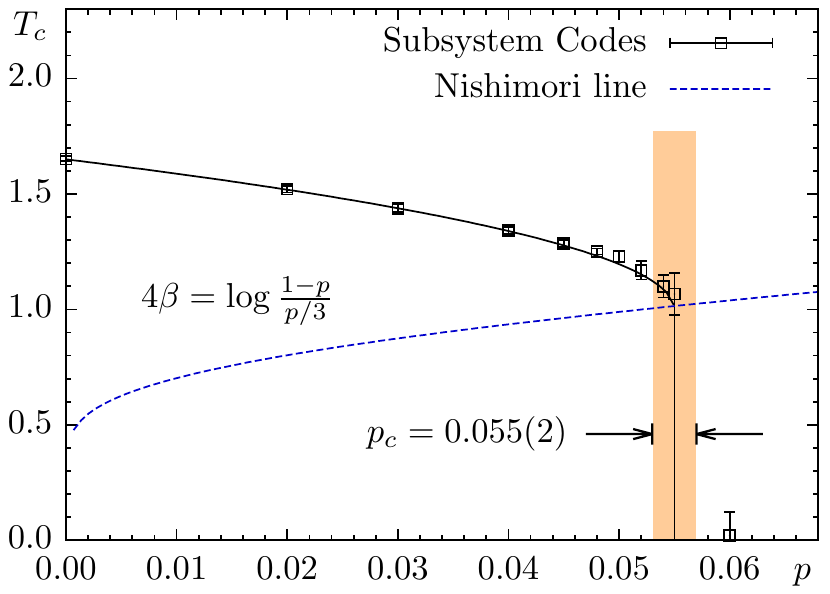}
    \end{minipage} \hfill \begin{minipage}[b]{3in}
	\caption{
	    Disorder--temperature phase diagram for subsystem codes
	    under depolarizing noise. The error threshold is given by
	    the intersection of the phase boundary (solid line) with the
	    Nishimori line (dashed). Our (conservative) numerical
	    estimate is $\pc=0.055(2)$ (orange vertical bar).  Despite
	    the low threshold compared to other proposals, this code is
	    very promising due to the simpler stabilizers involved in
	    the error-correction process.
	} \label{fig:subsystem:phasediagram}
    \end{minipage}
\end{figure}

Using Monte Carlo simulations, we compute the temperature--disorder
phase diagram for the aforementioned statistical-mechanical model (see
figure~\ref{fig:subsystem:phasediagram}) and estimate an error threshold
of $\pc=0.055(2)$ \cite{andrist:12}, which is remarkable given the
simplicity of the error-correction procedure.  Note that this
critical error rate is (numerically) smaller than the threshold
calculated for the toric code, as well as topological colour codes.  This
is a consequence of a compromise for a much simpler syndrome-measurement
and error-correction procedure: with a streamlined syndrome measurement
process, the physical qubits are given less time to decohere and the
error rate between rounds of error correction in an actual physical
realization will be smaller.

\subsection{Topological codes with measurement errors}

So far we have only considered different types of errors that might
occur on the physical qubits (at a rate $p$), while the process of
syndrome measurement was assumed to be flawless.  However, if additional
measurement errors occur (at a rate $q$), we need to devise a scheme
that can preserve quantum information over time even if intermediate
measurements are faulty.  This leads to the notion of so-called
``fault-tolerant'' codes: in this case, our best option is actually to
continuously measure and correct errors as they are detected. Note that
this introduces errors whenever the syndrome is faulty and these pseudo
errors can only be detected and corrected in a later round of error
correction.

This process of alternating measurement and correction phases can be
modeled by considering vertically stacked copies of the original quantum
setup, each representing one round of error correction. In this
simplified scenario, errors only occur at discrete time steps, and it is
instructive to think of the additional vertical dimension as a
\emph{time}. In particular, the measurements are represented by vertical
connections between the plaquettes where the corresponding stabilizer
resides.  Then, the state of each layer is related to the layer
immediately before it by the effect of the error channel, followed by
one round of syndrome measurement and error correction.  For no
measurement errors (i.e., $q=0$) all errors are detected perfectly and
are corrected within one step. Consequently, there is no inter-relation
between the errors found in consecutive layers. If $q>0$, however, some
errors can remain and new ones might be introduced due to the faulty
syndrome measured. In analogy to the error chains seen earlier, we refer
to these errors persevering over time as ``error histories.''

\begin{figure}
    \begin{minipage}[b]{4.1in}
	\includegraphics[width=\columnwidth]{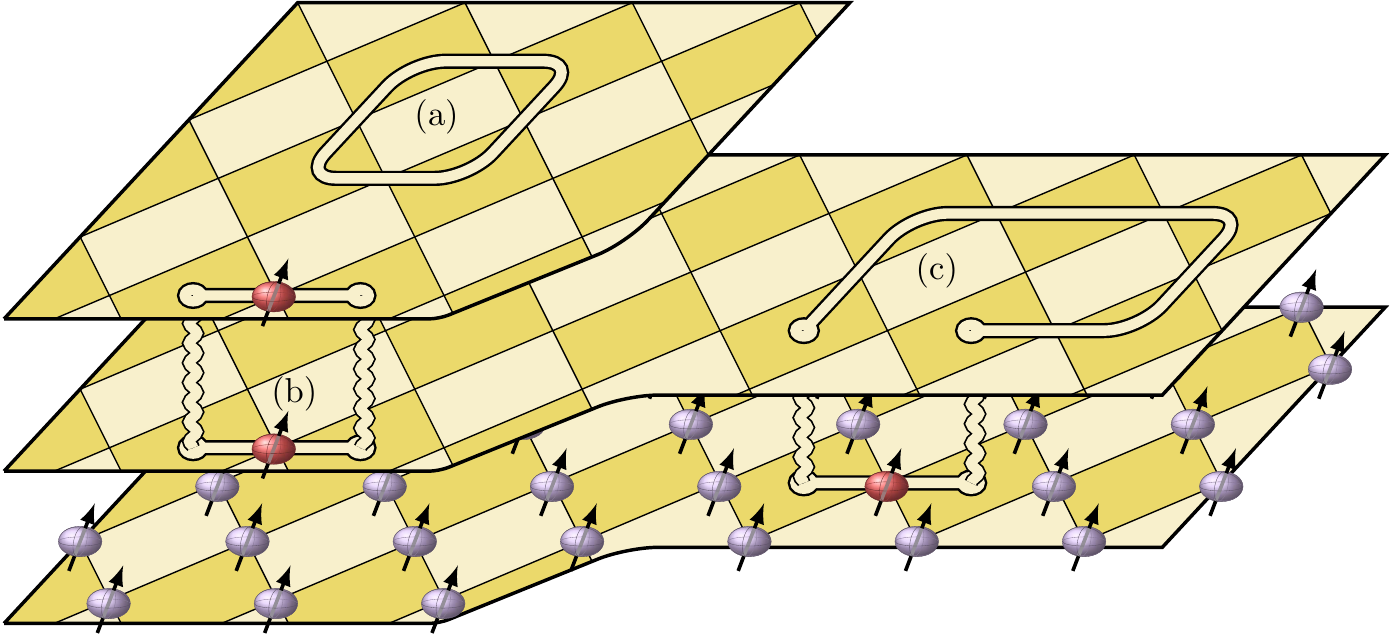}
    \end{minipage} \hfill \begin{minipage}[b]{2.1in}
	\caption{
	    Toric code with measurement errors. (a) Elementary loops are
	    error chains as seen before (figure~\ref{fig:toricspins}).
	    (b) Vertical links represent stabilizer measurements.  This
	    minimal error history consists of two consecutive qubit
	    flips that remain unnoticed due to faulty measurements.  (c)
	    A more complex error history.
		}
	\label{fig:gaugetoric}
    \end{minipage}
\end{figure}

Mapping the qubit-flip and measurement problem in the toric code to a
statistical-mechanical Ising model to compute the error threshold
\cite{dennis:02,wang:03,ohno:04} yields a Hamiltonian of them form
\begin{equation}
    \hamiltonian = 
        -J\sum_{\square} \tau_\square^{\ts{s}}
        \spinS_i \spinS_j \spinT_k \spinT_\ell
        -K\sum_{\square} \tau_\square^{\ts{t}}
        \spinT_i \spinT_j \spinT_k \spinT_\ell\,,
    \label{eq:gauge:hamiltonian}
\end{equation}
where the first [second] sum is over all qubits [vertical links] in the
lattice. Furthermore, we have introduced positive interaction constants
$J$ and $K$ to be chosen according to the (adapted) Nishimori
conditions~\cite{nishimori:81} $\exp(-2\beta J) = p/(1-p)$ and
$\exp(-2\beta K) = q/(1-q)$.  Note that each of the spatial spins
$\spinS$ represents an in-plane (i.e., horizontal) elementary loop that
consists purely of qubit flip errors, while the time-like spins $\spinT$
represent minimal error histories (vertical loops) that consist of two
qubit flip errors and two faulty measurements (see
figure~\ref{fig:gaugetoric}). As for the toric code without measurement
errors, these loops are used to tile the difference between two error
chains. Thus different spin configurations represent error chains of the
same class (sharing the same end points), albeit with different qubit
and measurement errors. And, given the Nishimori condition, the
Hamiltonian ensures that the Boltzmann weight corresponds to the relative
probability of each scenario.

Equation \eqref{eq:gauge:hamiltonian} describes a disordered Ising
lattice gauge theory with multi-spin interactions and four parameters,
$\beta J$, $\beta K$, $p$ and $q$. The mapping is valid along the
two-dimensional Nishimori sheet. We have treated spatial and time-like
equivalences separately to allow for different qubit and measurement
error rates. Interestingly, for the special case $p=q$, the resulting
Hamiltonian is isotropic, i.e.,
\begin{equation}
    \hamiltonian = \sum_\square \tau_{\smallsquare}
        \spin_i\spin_j\spin_k\spin_l\,,
    \label{eq:gauge:hamiltonian:pq}
\end{equation}
where $\square$ represents a sum over \emph{all plaquettes} in the lattice
(both vertical and horizontal). Because $p=q$ also implies $J=K$ via the
Nishimori condition, the model to investigate for this special case has only
two parameters, namely $\beta J$ and $p$.

\begin{figure}
    \begin{minipage}[b]{4.1in}
	\includegraphics[width=\columnwidth]{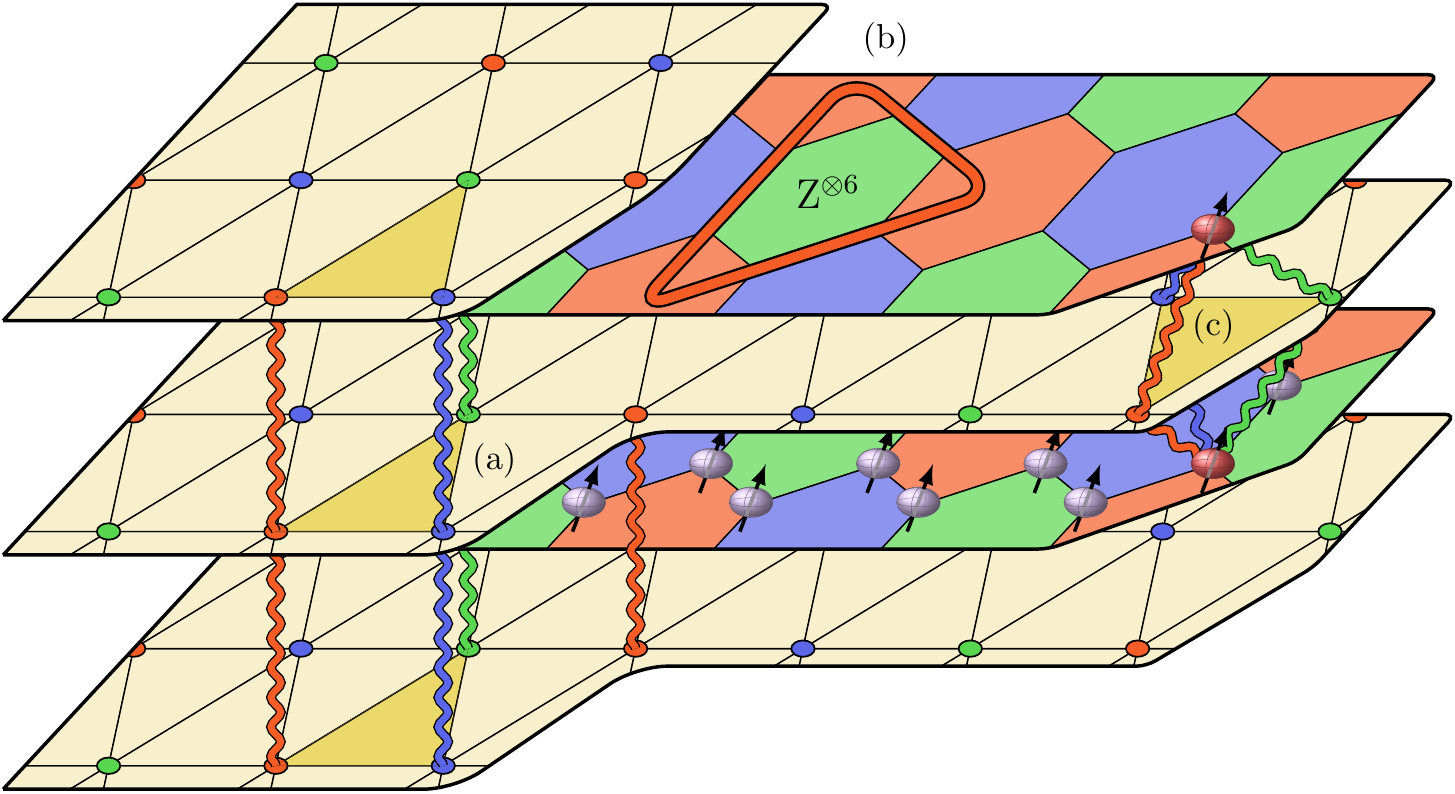}
    \end{minipage} \hfill \begin{minipage}[b]{2.1in}
	\caption{
	    Lattice structure for colour codes that consists of stacked
	    triangular and hexagonal layers.  (a) Vertical connections
	    represent the measurement history, while the qubits reside
	    on hexagonal lattice sites. The lattice gauge theory has two
	    elementary equivalences: (b) Coloured loops in the hexagonal
	    planes.  (c) Temporal equivalences consisting of two
	    consecutive qubit errors and three faulty measurements.
	} \label{fig:gaugecolor}
    \end{minipage}
\end{figure}

The model is a ${\mathbb Z}_2$ lattice gauge theory
\cite{wang:03,ohno:04}, so we cannot use a local order parameter to
determine the phase transition in our numerical simulations. Instead we
consider the peak in the specific heat and the distribution of Wilson
loop values to identify ordering in the system \cite{andrist:10}.  The
latter observable is interesting because the first-order transition
present in this system causes a double-peak structure near the
transition. Even though the effect is smeared out when disorder is
introduced, we can still reliably detect this shift of weight from one
peak to the other by performing a finite-size scaling analysis of the
skewness (third-order moment of the distribution). The temperature where
the skewness is zero represents the point where the distribution of
Wilson loops is double-peaked and symmetric, i.e., the phase transition.
Comparisons to more traditional methods, such as a Maxwell construction,
show perfect agreement. Note that this approach is generic and can be
applied to any Hamiltonian that has a first-order transition.

An analogous mapping and analysis is also possible for topological colour
codes with measurement errors.  In this case, one considers a
three-dimensional lattice consisting of stacked triangular layers, each
representing one round of error syndrome measurement.  The qubits reside
on intermediate hexagonal layers and are connected to their respective
check operators via vertical links as indicated in
figure~\ref{fig:gaugecolor}. In this case, the mapping to compute the
error threshold for faulty measurements and qubit flips produces a
${\mathbb Z}_2 \times {\mathbb Z}_2 $ lattice gauge theory given by the
Hamiltonian
\begin{equation}
    \hamiltonian = 
        -J \sum_{\tetra}\tau_{\tetra} 
            \spinS_i\spinS_j\spinS_k\spinT_l\spinT_m
        -K \sum_{\smallhex}\tau{\smallhex}
            \spinT_i\spinT_j\spinT_k\spinT_l\spinT_m\spinT_n\,,
    \label{eq:gauge:hamiltonian:color}
\end{equation}
where the first [second] sum is, again, over all qubits [measurements],
represented by the product of the five [six] equivalences affected by the
corresponding errors. The choice of the constants $J$ and $K$ is done as for
the toric code following the modified Nishimori conditions.  This Hamiltonian
also describes an Ising lattice gauge theory, but even for the choice of $p=q$
it is \emph{not} isotropic as was the case for the toric code with measurement
errors.

\begin{figure}
    \begin{minipage}[b]{3in}
	\includegraphics[width=3.0in]{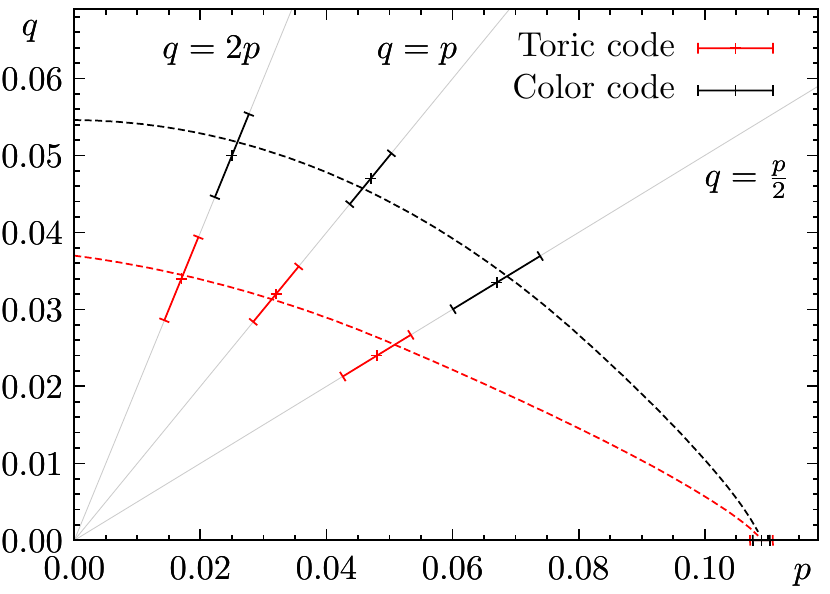}
    \end{minipage} \hfill \begin{minipage}[b]{3.0in}
	\caption{Preliminary results: Error threshold for bit flip
	errors (probability $p$) with different probabilities of
	measurement errors $q$ for the toric code (red) and colour codes
	(black).  For perfect error syndrome measurement, i.e., $q=0$
	(horizontal axis), the error thresholds of both codes agree,
	[$\pc=0.109(2)$].  However, for a non-vanishing measurement
	error rate ($q>0$), the error thresholds differ.  The dashed
	lines are guides to the eye. Numerical values listed in
	Table~\ref{tab:results}.} \label{fig:meas:phasediagram}
    \end{minipage}
\end{figure}

Both models for topological codes with measurement errors are studied
numerically using Monte Carlo simulations \cite{andrist:10}.
Interestingly, the thresholds calculated for the toric code and
topological colour codes do \emph{not} agree when possible measurement
errors are taken into account. While the toric code can only correct up
to $3.3\%$ errors when $p = q$, colour codes remain stable up to $4.8\%$.
This remarkable discrepancy is also seen in numerical studies where we
allow $p$ and $q$ to be different (as long as $q>0$), see
figure~\ref{fig:meas:phasediagram}.                       

\section{Summary and conclusions}
\label{sec:summary}

In Table \ref{tab:results} we summarize our results for different
combinations of topological codes and error sources. As can be seen, the
different proposals for topologically-protected quantum computation are
very resilient to the different error sources.

Note that the results for different error channels in Table
\ref{tab:results} cannot be compared directly: For the qubit-flip
channel, $\pc$ only refers to the maximum amount of flip errors that can
be sustained, while for the depolarizing channel $\pc$ is the sum of all
three basic error types. Furthermore, the lower threshold for subsystem
codes is the result of a compromise for a simpler error-correction
procedure. Likewise, a lower value of $\pc$ is to be expected in
fault-tolerant schemes due to the additional presence of measurement
errors. Remarkably, the error stability of the toric code and
topological colour codes appears to be different only in the
fault-tolerant regime where the mapping to a statistical-mechanical
model produces a lattice gauge theory, despite finding perfect agreement
in all other error channels.

We have outlined the mapping and subsequent analysis of several
topological error-correction codes to classical statistical-mechanical
Ising spin models with disorder. Because error chains correspond to
domain walls under this mapping, an ordered state in the classical model
can be identified with the scenario of feasible error correction, while
the proliferation of errors in the quantum setup is associated with a
disordered state in the classical model. After numerically calculating
the disorder--temperature phase diagram of the (classical) statistical
spin models, the error threshold can be identified with the intersection
point of the phase boundary with the Nishimori line. This critical error
threshold represents the maximum amount of perturbation each setup can
sustain and does not include the effects of realistic device
implementations.  However, the fact that these ``theoretical'' best-case
thresholds are so high is rather promising.

We conclude by highlighting again the beautiful synergy between quantum
error correction and novel disordered spin models in statistical
physics. We hope that these results will encourage scientists
specialized in analytical studies of disordered systems to tackle these
simple yet sophisticated Hamiltonians.

\begin{table}[H]
    \center
    \begin{tabular*}{0.92\textwidth}{@{\extracolsep{\fill}} l l l}
        \hline
        \hline
        Error Source&Topological Error Code&Threshold $\pc$\\ 
        \hline
        Qubit-Flip & Toric code		                       & $0.109(2)$\\
        Qubit-Flip & Colour code (honeycomb lattice)           & $0.109(2)$\\
        Qubit-Flip & Colour code (square-octagonal lattice)    & $0.109(2)$\\ \\
        Depolarization	& Toric code			       & $0.189(2)$\\
        Depolarization	& Colour code (honeycomb lattice)      & $0.189(2)$\\
        Depolarization	& Subsystem code     		       & $0.055(2)$\\ \\
        Qubit-Flip \& Measurement & Toric code		       & $0.033(5)$\\
        Qubit-Flip \& Measurement & Toric code ($q=2p$)	       & $0.017(3)$\\
        Qubit-Flip \& Measurement & Toric code ($q=p/2$)       & $0.047(6)$\\ \\
        Qubit-Flip \& Measurement & Colour code ($p = q$)      & $0.048(3)$\\
        Qubit-Flip \& Measurement & Colour code ($q=2p$)       & $0.025(4)$\\
        Qubit-Flip \& Measurement & Colour code ($q=p/2$)      & $0.066(7)$\\
        \hline
        \hline
    \end{tabular*}
    \caption{
    Summary of error thresholds calculated numerically. Note that our
    estimate for the toric code with qubit-flip and measurement errors
    agrees with the results of Wang \emph{et~al.}~\cite{wang:03} and
    Ohno~\emph{et~al.}~\cite{ohno:04}. The most precise estimate for the
    toric code with qubit-flip errors is $\pc = 0.10919(7)$, see
    \cite{parisen:09}.
	\label{tab:results}
    }
\end{table}

\ack{

Part of the figures shown in this work are from the PhD Thesis of Ruben
S.~Andrist (Eidgen\"ossische Technische Hochschule Diss.~No.~20588).
Work done in collaboration with H.~Bombin (Perimeter Institute),
M.~A.~Martin-Delgado (Universidad Complutense de Madrid), and
C.~K.~Thomas (Texas A\&M University). H.G.K.~acknowledges support from
the SNF (Grant No.~PP002-114713) and the NSF (Grant No.~DMR-1151387) and
would like to thank ETH Zurich for CPU time on the Brutus cluster and
Texas A\&M University for CPU time on the Eos, as well as Lonestsar
clusters.

}

\section*{References}

\bibliographystyle{iopart-num}
\bibliography{refs}

\end{document}